\documentclass[apjl]{emulateapj}
\usepackage{bm}
\usepackage{subfigure}
\usepackage{color}

\DeclareBoldMathCommand{\bV}{V}
\DeclareBoldMathCommand{\bv}{v}
\DeclareBoldMathCommand{\bu}{u}
\DeclareBoldMathCommand{\bx}{x}
\DeclareBoldMathCommand{\by}{y}
\DeclareBoldMathCommand{\bz}{z}
\DeclareBoldMathCommand{\br}{r}
\DeclareBoldMathCommand{\bb}{b}
\DeclareBoldMathCommand{\be}{e}
\DeclareBoldMathCommand{\bB}{B}
\DeclareBoldMathCommand{\bE}{E}
\DeclareBoldMathCommand{\bk}{k}
\DeclareBoldMathCommand{\bA}{A}
\DeclareBoldMathCommand{\bJ}{J}

\newcommand{\V}[1]{\mathbf{#1}} 
\newcommand{\T}[1]{\texttt{#1}}

\shorttitle{Current Sheets and Dissipation in Plasma Turbulence}
\shortauthors{TenBarge and Howes}

\begin{document}

\title{Current Sheets and Collisionless Damping in Kinetic Plasma Turbulence}

\author{J.~M. TenBarge}
\email{jason-tenbarge@uiowa.edu}
\affil{Department of Physics and Astronomy, University of Iowa, Iowa City, IA 52242, USA}
\author{G.~G. Howes}
\affil{Department of Physics and Astronomy, University of Iowa, Iowa City, IA 52242, USA.}

\begin{abstract}
We present the first study of the formation and dissipation of current
sheets at electron scales in a wave-driven, weakly collisional, 3D
kinetic turbulence simulation. We investigate the relative importance
of dissipation associated with collisionless damping via resonant
wave-particle interactions versus dissipation in small-scale current
sheets in weakly collisional plasma turbulence. Current sheets form
self-consistently from the wave-driven turbulence, and their filling
fraction is well correlated to the electron heating rate. However, the
weakly collisional nature of the simulation necessarily implies that
the current sheets are not significantly dissipated via Ohmic
dissipation. Rather, collisionless damping via the Landau resonance
with the electrons is sufficient to account for the measured heating
as a function of scale in the simulation, without the need for
significant Ohmic dissipation. This finding suggests the possibility
that the dissipation of the current sheets is governed by resonant
wave-particle interactions and that the locations of current sheets
correspond spatially to regions of enhanced heating.

\end{abstract}


\section{Introduction}
Turbulence plays an important role in space and astrophysical plasmas
by mediating the transfer of energy from large-scale motions to the
small scales at which the turbulence can be dissipated.  A major
unsolved problem is the identification of the physical mechanisms that
dissipate the small-scale turbulent motions, ultimately converting the
turbulent energy to plasma heat. The dynamics at the dissipative
scales are typically weakly collisional in diffuse astrophysical
plasmas, such as the solar wind, so the mechanisms responsible for the
dissipation and plasma heating are described by kinetic plasma
physics.  Two mechanisms have been proposed to be the dominantly
involved in the dissipation process for plasma turbulence:
collisionless wave-particle interactions
\citep{Howes:2008b,Schekochihin:2009,TenBarge:2012c} and  
dissipation in small-scale current sheets
\citep{Dmitruk:2004,Markovskii:2011,Matthaeus:2011,Osman:2011,Servidio:2011,Wan:2012a,Karimabadi:2013}.

In weakly collisional plasmas, it is well known that wave-particle
interactions lead to significant collisionless damping of the linear
kinetic wave modes. In
turbulent astrophysical plasmas, it has been proposed that the
fluctuations at perpendicular scales smaller than the ion Larmor
radius, $k_\perp\rho_i \gtrsim 1$, have properties typical of kinetic
Alfv\'{e}n waves
\citep{Howes:2008a,Howes:2008b,Howes:2008c,Schekochihin:2009,Salem:2012},
and will therefore suffer collisionless damping.  In the case of the
solar wind, it has been suggested
\citep{Howes:2008b,Schekochihin:2009,Howes:2011c,Howes:2011b} that
electron Landau damping dominates the dissipation of these turbulent
electromagnetic fluctuations at $k_\perp\rho_i \gtrsim 1$. Free energy
transferred conservatively to the particle distribution functions by
wave-particle interactions is ultimately thermalized by arbitrarily
weak collisions through the action of an entropy cascade in phase
space \citep{Schekochihin:2009}.  

On the other hand, a number of recent studies focusing on the
intermittent structures that inherently develop in plasma
turbulence have suggested that dissipation dominantly occurs in
coherent structures, in particular, small-scale current sheets
\citep{Dmitruk:2004,Markovskii:2011,Matthaeus:2011,Osman:2011,Servidio:2011,Wan:2012a,Karimabadi:2013}. The kinetic physical mechanism by which dissipation occurs in current
sheets has not been clearly elucidated. Hybrid kinetic-ion and fluid-electron
simulations in 2D suggest stochastic perpendicular ion heating due to demagnetization in current sheets \citep{Parashar:2009,Markovskii:2011}, 2D and 3D Particle-in-Cell (PIC) simulations of reconnection suggest the acceleration of electrons by
parallel electric fields \citep{Drake:2003,Pritchett:2004,Egedal:2008,Egedal:2009,Egedal:2010,Egedal:2012} and/or Fermi acceleration \citep{Drake:2006}, and 2D gyrokinetic simulations suggest linear phase mixing/Landau damping \citep{Loureiro:2013}. Temperature measurements in the near-Earth solar wind have been used to both support \citep{Osman:2011,Osman:2011a} and refute \citep{Borovsky:2011} the
proposal that plasma heating dominantly occurs in current sheets.

In this Letter, we present a wave-driven, 3D gyrokinetic turbulence
simulation at scales smaller than the ion Larmor radius that
self-consistently generates small-scale current sheets. We find that
the current sheet filling fraction is well correlated with the
electron heating rate in the simulation. Yet, Ohmic dissipation is
negligible, and the measured electron heating rate by scale is well
reproduced by assuming dissipation is entirely associated with
electron Landau damping, suggesting the possibility that the current
sheets are damped collisionlessly by resonant wave-particle
interactions and correspond to regions of local heating.

\section{Kinetic Simulation}\label{sec:kin_sim}
The simulation 
was performed with the Astrophysical Gyrokinetics code,  \T{AstroGK}
\citep{Numata:2010}, which solves the equations of gyrokinetics
\citep{Frieman:1982,Howes:2006}. Collisions are treated using a fully
conservative, linearized, and gyroaveraged collision operator
\citep{Abel:2008,Barnes:2009}. The simulations are driven at the
simulation domain scale with an oscillating Langevin antenna coupled
to the component of the vector potential parallel to the equilibrium
magnetic field, $\V{B}_0=B_0 \hat{\V{z}}$ \citep{TenBarge:2012b}.

The simulation models a proton and electron plasma with a realistic
mass ratio $m_i/m_e = 1836$, $\beta_i =1$, and $T_i/T_e =1$, where
$\beta_i =v_{ti}^2/v_A^2$, $v_A$ is the
Alfv\'{e}n speed, and $v_{ti} = \sqrt{2T_i/m_i}$ is the ion thermal
speed. The simulation employs a periodic domain of size $L_\perp^2
\times L_z$, elongated along the straight, uniform equilibrium magnetic
field $\V{B}_0$. Relevant parameters are $k_\perp
\rho_i \in [5, 105]$, $k_z \rho_i/\epsilon \in [1,16]$, $A_0/\epsilon
\rho_i B_0 = 0.2$, $\nu_i \rho_i / v_{ti} \epsilon = 0.2$, and $\nu_e
\rho_i / v_{ti} \epsilon = 0.5$, where $\rho_i = v_{ti}/ \Omega_i$ is
the ion Larmor radius, $\Omega_i$ is the ion
gyrofrequency, $\epsilon = 2 \pi \rho_i / L_z \ll 1$ is the
gyrokinetic expansion parameter, $A_0$ is the antenna amplitude, and
$\nu_s$ is the collision frequency of species $s$. Time is normalized to
the linear frequency of a kinetic Alfv\'{e}n wave at
the simulation domain scale, $\omega_0 = 3.6k_{z0} v_A = 3.6
\omega_{A0}$. Therefore, the corresponding domain scale turn-around
time is $\tau_0 = 2\pi/\omega_0 \simeq
1.75\omega_{A0}^{-1}$. Collision frequencies are chosen to prevent
build-up of small-scale structure in velocity space but remain small
enough  not to alter the weakly collisional dynamics: $\nu_s\ll
\omega_0$ is satisfied, so the simulation is weakly collisional. A value for $\epsilon \sim \delta B / B_0$ can be estimated by examining solar wind magnetic energy spectra at our simulation domain scale, $k \rho_i = 5$. Based on spectra available in \citet{Alexandrova:2009,Sahraoui:2009,Sahraoui:2010b}, $\epsilon \sim \delta B / B_0 \sim 0.01$ at $k \rho_i = 5$ in the solar wind. The antenna amplitude is chosen to satisfy critical balance at the domain scale,
so the simulation represents critically balanced, strong
turbulence. Analysis of a similar simulation \citep{TenBarge:2012c}
demonstrates a magnetic energy spectrum in excellent agreement with
the empirical form found from a large statistical sample of
dissipation range measurements in the solar wind
\citep{Alexandrova:2012}.

\begin{figure}[top]
\subfigure{
\includegraphics[width=\linewidth]{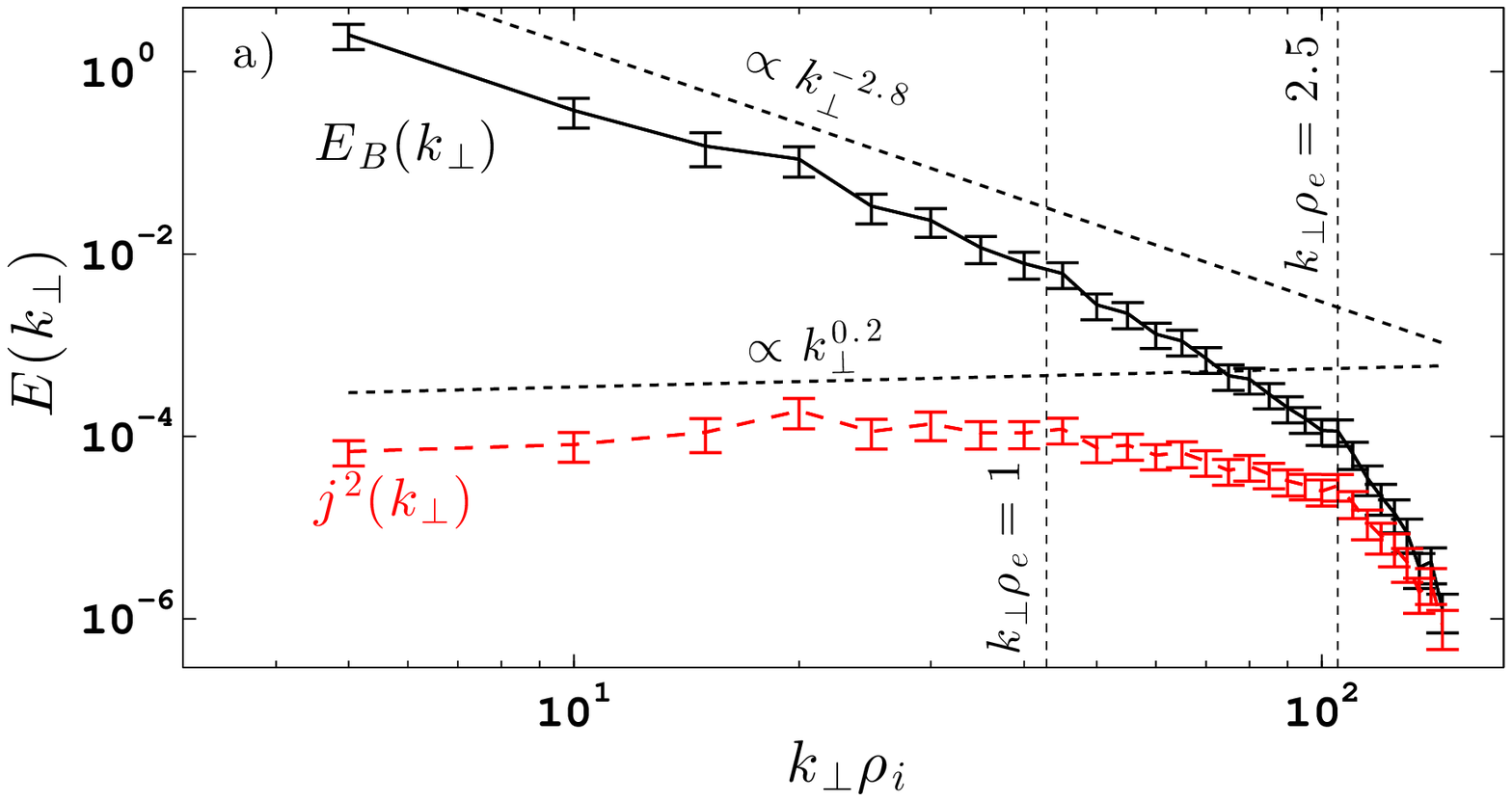}}
\subfigure{
\includegraphics[width=\linewidth]{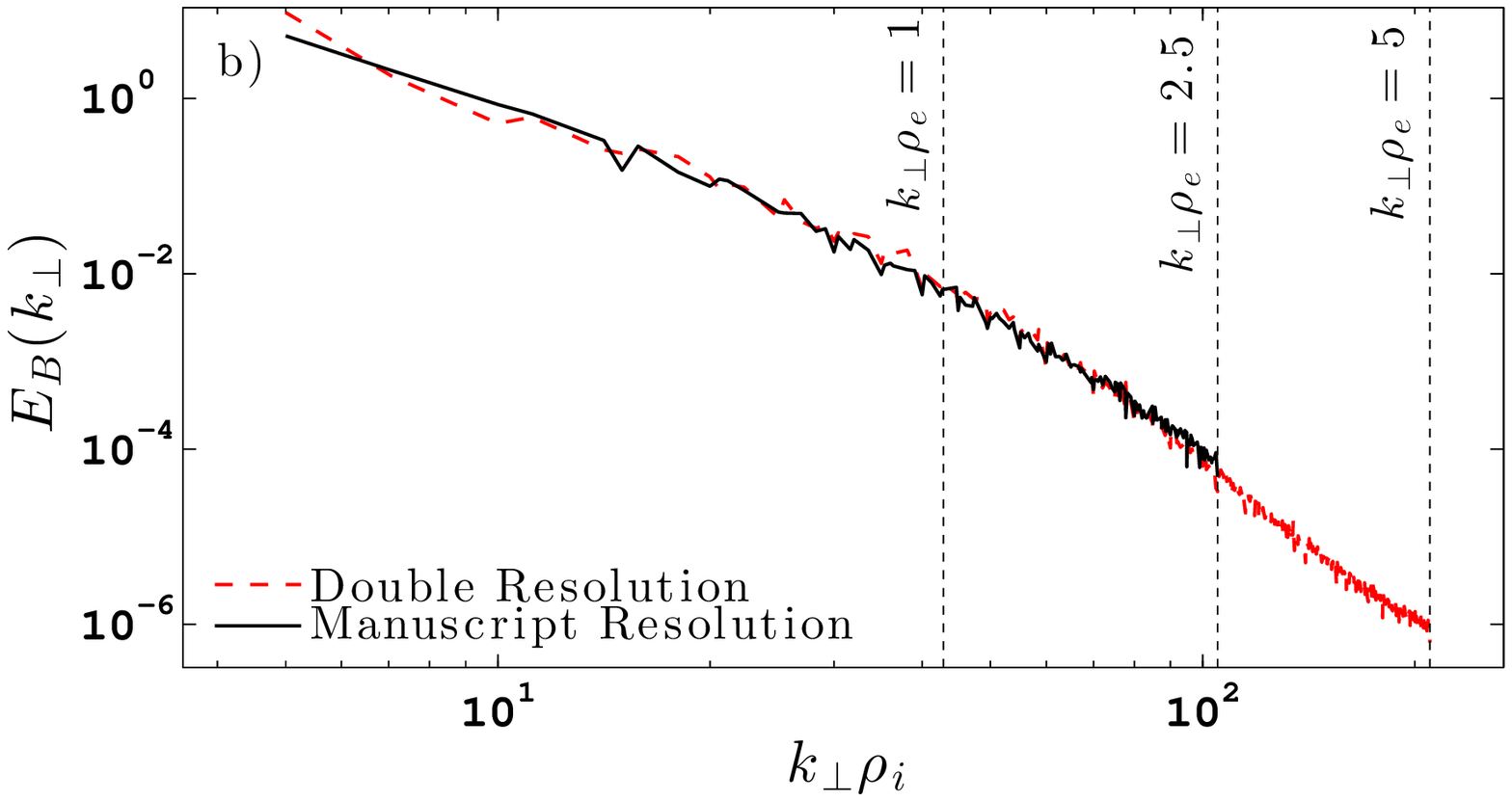}}
\caption{a) Time averaged one-dimensional trace magnetic 
energy spectrum (solid) and the spectrum of the square of the current density
(dashed). b) Instantaneous one-dimensional trace magnetic energy spectra for the same resolution as the manuscript (solid) and double resolution (dashed), demonstrating numerical convergence.}
\label{fig:current}
\end{figure}

\section{Magnetic Energy Spectrum and Current Density Spectrum}

First we demonstrate that the magnetic energy spectrum from our
simulation is consistent with measurements in the solar wind and then
present the perpendicular wavenumber spectrum of the square of the current
density, $j^2(k_\perp)$. In Figure \ref{fig:current}a), we plot (solid)
the average one-dimensional trace magnetic energy spectrum
$E_B(k_\perp)$, where the total magnetic energy $E_{B}^{(tot)}= \int
dk_\perp E_{B}(k_\perp)$.  The average is performed over the
steady-state evolution of the system, $ 1.5 \tau_0 \le t \le 4.1
\tau_0$; error bars represent the variance over the same interval.
This spectrum is quantitatively consistent with a large sample of
measurements of the dissipation range magnetic energy spectrum in the
solar wind \citep{Alexandrova:2012}, suggesting that this simulation
contains the essential physical ingredients underlying turbulence in
the solar wind. Since $j = |\V{j}|= |(c/4\pi) \nabla \times \V{B}|$,
we expect to find $j \propto kB$, and therefore the scaling of
$j^2(k_\perp) = \int k_\perp d\phi d k_\parallel j^2(\V{k})$ should
satisfy the relation $j^2(k_\perp) \propto k_\perp^3 E_{B}(k_\perp)$,
as confirmed by the plotted spectrum (dashed) of $j^2(k_\perp)$.

\begin{figure}[top]
\includegraphics[width=\linewidth]{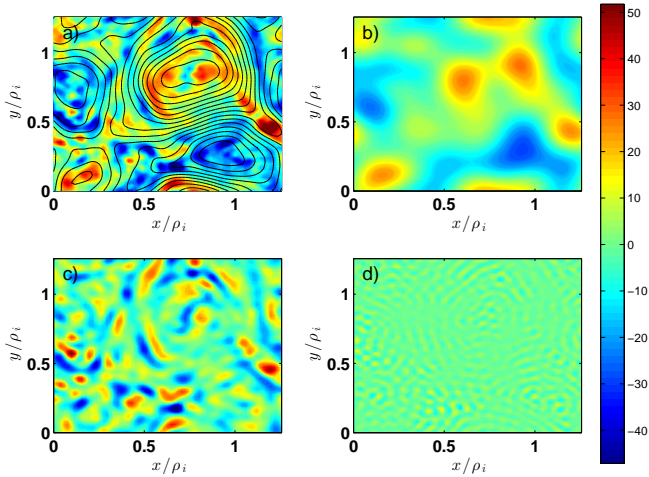}
\caption{Parallel current density, $j_z$, for a 
perpendicular plane at $t=2.43 \tau_0$, with different band-pass
filters applied: (a)~unfiltered, (b)~$5 \le k_\perp \rho_i < 21$,
(c)~$21 \le k_\perp
\rho_i < 84$, and (d)~$k_\perp \rho_i \ge 84$.  Contours of the parallel vector potential $A_z$
are shown in (a).}\label{fig:sheets}
\end{figure}

\section{Current Sheet Formation}
In a weakly collisional plasma with a guide magnetic field, current sheets are not expected to form at scales below $k_\perp \rho_e \sim 1$. In Figure \ref{fig:sheets}, we plot (a) the parallel current density $j_z(x,y)$ from a perpendicular cut through
the simulation domain. To explore the contribution to the current from
different scales, we present spatially band-pass filtered data: (b)~$5
\le k_\perp \rho_i < 21$, (c)~$21 \le k_\perp \rho_i < 84$, and (d)~$k_\perp \rho_i \ge 84$. The large-scale currents visible in panels
(a) and (b) are dominated by the driving, which generates upward and
downward propagating kinetic Alfv\'{e}n waves with $k_\perp \rho_i =
5$. Panel (c), whose filter is approximately centered on the electron
gyroradius, $0.49 \le k_\perp \rho_e \le 1.95$, shows that this 3D
gyrokinetic simulation indeed produces current sheets at the $\rho_e$
scale, consistent with such development in a wide range of plasma
turbulence simulations. These electron-scale diffusion regions are
highly intermittent, both spatially and temporally, with a typical
lifetime $\tau \lesssim 0.1 \tau_0$. The lack of significant current
density in panel (d) shows that current sheets do not form at scales
$k_\perp \rho_e > 2$. To confirm that our simulation indeed
has sufficient perpendicular resolution to capture the current sheet
dynamics, we ran a convergence test with double the resolution in the
perpendicular plane. Plotted in Figure \ref{fig:current}b) are the instantaneous one-dimensional trace magnetic energy spectrum for the manuscript resolution (solid) extending to $k_\perp \rho_e = 2.5$ and the same spectrum after doubling the resolution (dashed) to $k_\perp \rho_e = 5$ and allowing the simulation to saturate, demonstrating that the magnetic energy spectrum is resolved. The spatially filtered
current density for the double resolution simulation shows similar
results to Figure~\ref{fig:sheets}d, that no current sheet structure
forms at scales $k_\perp \rho_e > 2$, confirming that our simulation
has sufficient perpendicular spatial resolution to capture fully the
electron-scale current sheet dynamics.


Note that we are using the general definition of current sheet as a discontinuity in the magnetic field. This is consistent with the definition of current sheet used in recent solar wind literature, e.g., \citet{Vasquez:2007,Greco:2009,Osman:2011}. We do not attempt to differentiate between current sheets associated with reconnecting
magnetic flux and those arising from interfering Alfv\'{e}n waves.

It is also important to note that the simulation is driven by
injecting domain-scale waves, which generate strong turbulence. The
current sheets form self-consistently from the cascade of wave-driven
turbulence and are not seeded or otherwise initialized to
form. Therefore, we conclude that electron-scale current sheets form
as a natural consequence of Alfv\'{e}nic turbulence in this 3D
gyrokinetic simulation.

\section{Current Sheets and Electron Heating}
We next examine the relative contribution to the measured electron heating from
wave-particle interactions and dissipation in current sheets. The
analytical equations for plasma heating
\citep{Howes:2006} have been implemented as a diagnostic in
\T{AstroGK}. Boltzmann's $H$ Theorem states that the entropy increase
necessary for irreversible heating requires collisions
\citep{Howes:2006}, so the collisional heating (plus a small amount of numerical dissipation) is used to measure the heating rate of each plasma species.  Over the range of scales simulated, $k_\perp \rho_i
\in [5, 105]$, little ion heating occurs, so we focus on the electron
heating. The collisional electron heating is given by
\begin{eqnarray}
Q_e& =&   -\sum_{\V{k}_\perp} \int^{L_z}_{-L_z} \frac{dz}{2 L_z} \int d^3 \V{v}
\frac{T_{0e}}{F_{0e}} 
\\ & & \left[ h_{e\V{k}_\perp}\left(
 \frac{\partial h^*_{e\V{k}_\perp}}{\partial t} \right)_{\mbox{coll}}
 +  h^*_{e\V{k}_\perp} \left(\frac{\partial  h_{e\V{k}_\perp}}{\partial t}
 \right)_{\mbox{coll}}\right], \nonumber
\end{eqnarray}
where $h_{e\V{k}_\perp}= h_e(k_x,k_y,z,v_\parallel,v_\perp,t)$ is the
non-Boltzmann portion of the perturbed electron distribution function and
$F_{0e}$ is the equilibrium electron distribution function
\citep{Howes:2006,Numata:2010}. The collisional electron heating as a
function of perpendicular wavenumber $Q_e (k_\perp)$ is computed by
summing over annular rings in the perpendicular plane such that the
total heating is given by $Q_e = \int d k_\perp Q_e (k_\perp)$. Note
that, in the weakly collisional limit, the heating rate is independent
of the collision frequency
\citep{Howes:2006}. For the steady-state evolution of the simulation
over $1.5 \tau_0 \le t \le 4.1 \tau_0$, the heating diagnostics
recover total power balance to $\lesssim 2\%$
\citep{TenBarge:2012c}.

\begin{figure}[top]
\includegraphics[width=\linewidth]{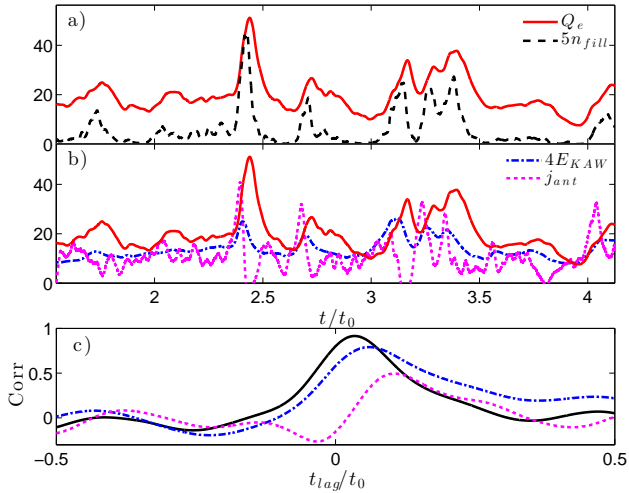}
\caption{a) Volume filling fraction of current sheets 
satisfying $j > j_{max}/3$ in percent (black dashed) and the electron collisional heating
rate $Q_e$ (red solid). b) The injected antenna power (magenta dotted) and the total turbulent energy, $E_{KAW}$, (blue dash-dotted). c) Cross correlations between the electron heating rate and filling fraction (black solid), antenna power (magenta dotted), and $E_{KAW}$ (blue dash-dotted).}\label{fig:heatvstime}
\end{figure}

To estimate the contribution of current sheet dissipation to the
heating rate, we compare the fraction of volume occupied by current
sheets to the electron heating rate as a function of time. The volume
filling fraction of current sheets is computed as the percentage of
the volume with current density $j> j_{th}$, with a chosen threshold
$j_{th} = j_{max}/3$, where $j_{max}$ is the maximum current density
over all time and space in the simulation. Varying this threshold alters the magnitude of the filling fraction but not the form of its variation with time. In Figure~\ref{fig:heatvstime}~a), we plot the volume filling fraction in percent (black dashed) and the electron collisional heating rate, discussed above, $Q_e$ (red solid)
as a function of time---all quantities in the figure have been
integrated over the entire simulation domain. In
Figure~\ref{fig:heatvstime}~b) is plotted the boxcar averaged (over
$\Delta t= 0.13 \tau_0$) power injected into the plasma by the
Langevin antenna (magenta dotted) and the total energy of the
turbulent fluctuations in the simulation including the magnetic field
and kinetic energies $E_{KAW} = E_{B_\perp} + E_{B_\parallel} +
E_{KE}$ (blue dash-dotted), where $E_{KE} =
\sum_s m_s n_{0s} u_s^2 /2$ and $\V{u}_s$ is the fluid velocity of each species. 

The cross correlations between the electron heating rate and filling
fraction (black solid), antenna power (magenta dotted), and $E_{KAW}$
(blue dash-dotted) are plotted in Figure~\ref{fig:heatvstime}~c). The
injected antenna power and the electron heating rate are poorly
correlated, $\langle\mbox{max}(\mbox{Corr}(Q,j_{ant}))\rangle = 0.52
\pm 0.02$---the mean and variance are calculated from the present
simulation and five other identically prepared simulations employing
different random number seeds for the turbulent driving. Similarly,
the electron heating rate and total turbulent energy are not well
correlated, $\langle\mbox{max}(\mbox{Corr}(Q,E_{KAW}))\rangle = 0.78
\pm 0.04$, suggesting that the heating rate is not simply a function
of the driving or magnitude of turbulent energy. On the other hand,
the electron heating rate is well correlated with the current sheet
filling fraction, $\langle\mbox{max}(\mbox{Corr}(Q,n_{fill}))\rangle =
0.91 \pm 0.04$. The strong correlation suggests that dissipation
associated with current sheets plays an important role in heating the
electrons.

\begin{figure}[top]
\includegraphics[width=\linewidth]{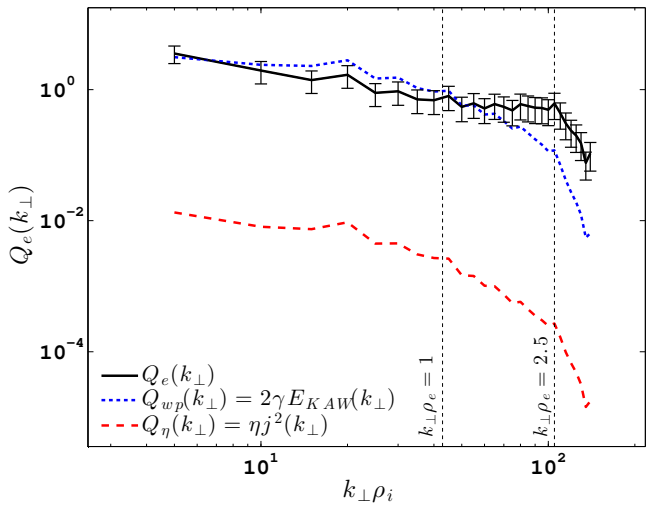}
\caption{Measured heating  of the electrons from the 
simulation by perpendicular wavenumber, $Q_e(k_\perp)$ (solid), an
estimate of the electron heating based on linear wave-particle damping
$Q_{wp}(k_\perp)$ (dotted), and the Ohmic heating rate
$Q_\eta(k_\perp)$ (dashed).}\label{fig:heating}
\end{figure}

\section{Electron Heating by Scale}
In Figure \ref{fig:heating}, we present a plot of the electron
collisional heating rate by perpendicular wavenumber, $Q_e(k_\perp)$
(solid), averaged over $1.5 \tau_0 \le t \le 4.1 \tau_0$, where error
bars represent the variance over the interval. The instantaneous shape of the heating curve is similar to the average. This plot shows that the electron heating is nearly constant over all scales, with about half of the total heating
occurring at scales $k_\perp \rho_e < 1$. The turn-down at $k_\perp
\rho_i > 105$ is an artifact due to the diminishing number of Fourier
modes in the corner beyond the fully resolved simulation domain.

As a function of $k_\perp$, we may predict the collisionless damping
of the turbulent fluctuations by resonant wave-particle interactions
in our simulation using $Q_{wp}(k_\perp) = 2 \gamma(k_\perp)
E_{KAW}(k_\perp)$, where $\gamma$ is the \emph{linear} kinetic damping rate
of kinetic Alfv\'{e}n waves (dominated by electron Landau
damping). This prediction for the wave-particle interaction heating
rate requires integration over $k_\parallel$, where parallel is with
respect to the local magnetic field and is typically determined via
structure functions or wavelets. To avoid the complications of
determining the local magnetic field direction, we use frequency as a proxy for
the parallel wave vector since $\omega \propto k_\parallel$ for
kinetic Alfv\'{e}n waves, \citet{TenBarge:2012a}. This prediction for wave-particle damping, plotted in Figure~\ref{fig:heating} (dotted), admits \emph{no free
parameters}, yet it agrees well with the measured collisional
electron heating (solid): the integrated, total predicted electron
heating is within $4\%$ of the collisional heating diagnosed in \T{AstroGK}. 
The slight excess of wave-particle damping at $5 < k_\perp \rho_i <
40$ and of electron collisional heating at $k_\perp \rho_i >
40$ is consistent with the action of the electron entropy cascade
\citep{Schekochihin:2009}. Through the entropy cascade, energy removed
by electron Landau damping at $5 < k_\perp \rho_i <40$ is expected to
appear as collisional heating at higher wavenumbers.


We also compute the Ohmic heating rate $Q_\eta = \eta j^2$, where
$\eta = 0.38(4\pi)\nu_{ei} d_e^2/c^2$ is the Spitzer resistivity
\citep{Spitzer:1953}, $\nu_{ei} = \nu_e$ is the electron-ion 
collision frequency, $d_e = c/\omega_{pe}$ is the electron inertial
length, and $\omega_{pe}$ is the electron plasma frequency.  The Ohmic
heating rate is plotted (dashed) in Figure~\ref{fig:heating}, clearly $Q_\eta \ll Q_{wp} \simeq Q_e$. Theory predicts negligible Ohmic heating for a weakly collisional plasma, since electron-ion collisions are insufficient to significantly heat the
electrons. Therefore, Ohmic dissipation of the current cannot account
for the observed electron heating in the simulation, despite the
strong correlation between current sheet filling fraction and heating
rate.

\section{Discussion}
A puzzling aspect of these results is that, despite the clear correlation between the electron heating rate and the volume filling fraction of current sheets, the electron
heating as a function of wavenumber is well predicted assuming that
Landau damping is entirely responsible for the electron heating.  This
unexpected agreement raises the interesting possibility that the
dissipation of the current sheets in the simulation occurs entirely via Landau
damping. 

The solution to this puzzle lies in the relationship between the
current and magnetic field, namely $j \propto k B$. This relationship
implies that regions of strong current correspond to regions with
enhanced small-scale magnetic structure since the current is weighted
toward small spatial scales. Since the linear kinetic damping rate
increases with wavenumber, regions with enhanced small-scale magnetic
structure will also correspond to regions of enhanced wave-particle
damping. Therefore, regions of strong current may also correspond to
regions of enhanced wave-particle damping.

\begin{figure}[top]
\includegraphics[width=\linewidth]{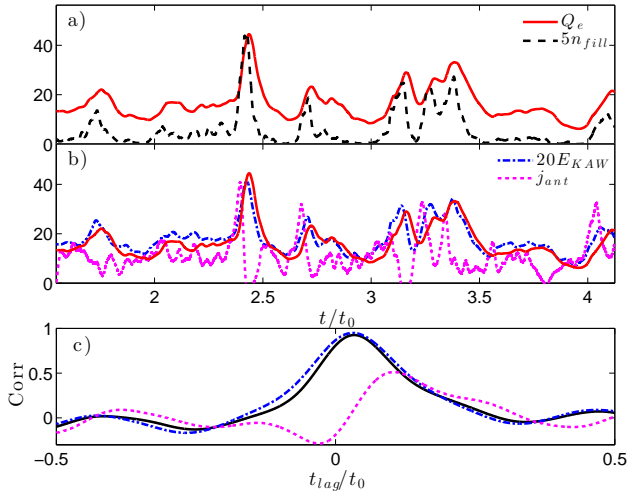}
\caption{Quantities as in Figure~\ref{fig:heatvstime} 
with a high-pass filter applied to remove modes with $k_\perp \rho_i <
10$, except the antenna current.}\label{fig:heatvstime_filt}
\end{figure}

To test this hypothesis, we apply a high-pass filter to the data
presented in Figure~\ref{fig:heatvstime} retaining only modes with
$k_\perp \rho_i \ge 10$---no filter is applied to the antenna
current. The result of the filtering is plotted in
Figure~\ref{fig:heatvstime_filt}, where all of the cross correlations
exceed $0.9$, with the exception of the antenna power, which remains
poorly correlated.

The picture suggested by this simulation is one in which current
sheets are self-consistently formed by the interaction of kinetic
Alfv\'{e}n wave-like fluctuations, where each of the fluctuations is
dissipating at its Landau damping rate. Therefore, current sheet
formation and dissipation is dominated by the evolution of the
Alfv\'{e}nic turbulence, and current sheets correspond to sites of
enhanced dissipation and heating, as suggested by recent analyses of
solar wind turbulence \citep{Osman:2011,Osman:2011a}. 

The validity of the Landau prescription described in this manuscript is predicated on the distribution function not deviating significantly from Maxwellian. This can be simply tested by examining the electron fluid velocity moments. We find that $\mbox{max}(u_{ze} / v_{te}) \sim 0.01$ based on the value of $\epsilon$ determined in \S \ref{sec:kin_sim}, suggesting that the Landau prescription is indeed valid. A more detailed examination of the electron distribution functions within regions of intense current in our simulations is necessary to confirm the dominance of the dissipation by electron Landau damping.



\section{Conclusion}
We find that electron scale current sheets develop self-consistently as a consequence of wave-drive turbulence in a 3D gyrokinetic simulation of the dissipation range, $k_\perp \rho_i \in [5, 105]$. The electron heating rate is well correlated with the
volume filling fraction of current sheets, suggesting that the
dissipation of current sheets plays an important role in the heating
of electrons. However, the electron heating rate as a function of
scale is well predicted by assuming that all dissipation is due to
collisionless damping of the turbulent fluctuations via Landau
resonance with the electrons. In the weakly collisional plasma, Ohmic
dissipation of current sheets is negligible. Due to the relationship
between the current and magnetic field, significant current highlights
regions with enhanced small-scale magnetic structure, which will be
collisionlessly damped at a rate greater than surrounding plasma. This
suggests that current sheets may correspond spatially to locations of
enhanced dissipation and heating, regardless of whether that
dissipation is due to collisionless wave-particle interactions or
active magnetic reconnection.


\acknowledgments
The authors thank Homa Karimabadi. This work
was supported by NASA grant NNX10AC91G and NSF CAREER grant
AGS-1054061, resources of the Oak Ridge Leadership
Computing Facility, supported by the Office of Science of the U.S. DOE
under Contract No.  DE-AC05-00OR22725, and advanced computing resources provided by the NSF, partly performed on Kraken at the National Institute for Computational Sciences.


\end{document}